# A Method to Determine the Ultrafast Laser Beams Spot Size

A. Srinivasa Rao[1,2*], Alok Sharan[1]

[1]*Department of Physics, Pondicherry University, RV Nagar, Kalapet, Puducherry, India-605014*
[2]*Photonics Sciences Lab., Physical Research Laboratory, Ahmedabad, Gujarat*
[*]*Email: sri.jsp7@gamil.com*

**Abstract:** From the standard $TEM_{00}$ Gaussian beam profile equations, we have derived the equations for beam waist at lens focus as a function of variable spot size. In this process, we obtained two equations for beam waist and the validity of these equations is studied with respect to Rayleigh length. The physical validity of the equations has been theoretically checked and also experimentally verified with He-Ne laser at 632 nm. The derived equations are useful for the estimation of the spot size of high peak power lasers (those spot sizes less than few mm). In complicated experimental setup by inserting a beam splitter, we can walk out the some of the beam and the spot size of this beam at the particular position can be measured by normal methods (pinhole and knife edge techniques). By using our derived equations we can know the spot size of the beam at any position in the experimental setup form the measured spot size.

## 1. Introduction

In the discussion of laser experimental results either through intensities or fluencies, it is imperative to know the spot size of the laser beam [1-4]. The spot size of the Gaussian laser beam can be measured by different techniques: burn spot method [5], slit [6], CCD camera [7], quadrant photodiode [8], knife-edge/pinhole [9-13] and periodic ruling techniques [14-19]. The periodic ruling technique is a fast technique but the fabrication of ruling is a difficult process. CCD camera, and quadrant photodiode techniques are expensive. The knife-edge and Pinhole are the conventional techniques for spot-size measurement. Pinhole technique can also be used for measuring intensity profile and the resolution depends on the size of the pinhole. Spot size determination of ultra-short laser pulses [20] by this technique is not advisable as high peak power can easily damage the detectors. To overcome this problem, we present here an alternative method for determining the spot sizes at any position.

## 2. Theoretical model

Normally the femto-second laser beams from the OPA have the spot sizes around 1-2 mm. Due to high peak power, it is not easy to measure small spot sizes. In the process of explanation, let consider a beam propagating along the *z*-direction. We can focus the incident laser pulse of 1-2 mm spot size and then choose a z position along the optic axis to measure the spot size such that intensity is sufficiently below the damage threshold intensity of the detector. Theoretically, we can get the beam waist in terms of the measured spot size and its position (by using Eqs.4 and 5). As a result, we can obtain the spot size at required position by using Eq. 1.

Here we have used the condition for the propagation of the Gaussian beam through the focus (Eqs. 1 and 2) to obtain the expression for beam waist in terms of variable spot size and its position. $\omega_0$ is the beam waist and $z_0$ is the Rayleigh length.

$$\omega(z) = \omega_0 \sqrt{1 + z^2/z_0^2} \qquad (1)$$

$$z_0 = \frac{\pi \omega_0^2}{\lambda} \qquad (2)$$

Eqns. 1 and 2 together gives the fourth order equation for $\omega_0$ in terms of the variable position of laser beam and corresponding spot size $\omega(z)$.

$$\omega_0^4 - \omega_0^2 \omega^2(z) + \frac{z^2 \lambda^2}{\pi^2} = 0 \qquad (3)$$

Eq. 3 has the following two physically reliable solutions out of four solutions.

$$\omega_0 = \frac{\omega(z)}{\sqrt{2}} \sqrt{\left(1 + \sqrt{1 - \left(\frac{2z\lambda}{\pi \omega^2(z)}\right)^2}\right)} \qquad (4)$$

$$\omega_0 = \frac{\omega(z)}{\sqrt{2}} \sqrt{\left(1 - \sqrt{1 - \left(\frac{2z\lambda}{\pi \omega^2(z)}\right)^2}\right)} \qquad (5)$$

Further, these two solutions are valid only when Eq. 6 has satisfied.

$$f(z) = z^4 + z_0^4 - 2z^2 z_0^2 \geq 0 \qquad (6)$$

Eq. 3 has been derived from well-known standard equation Eqs. 1 and 2. Therefore Eqs. 4 and 5 are intended to provide the physically acceptable solution for Eq. 3 with satisfying inequality 6. The beam waist should be a real valued number and given only when the inequality 6 satisfies.

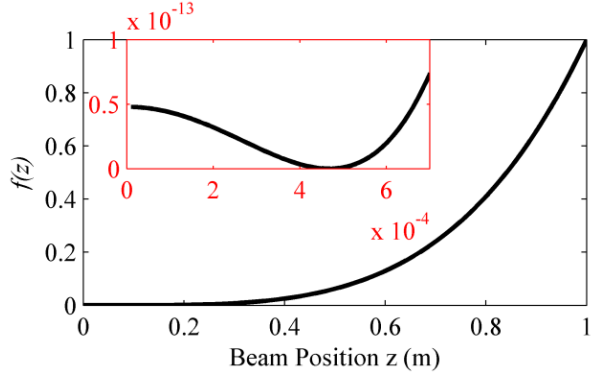

**Fig. 1** plot of inequality function *f(z)* as a function of z.

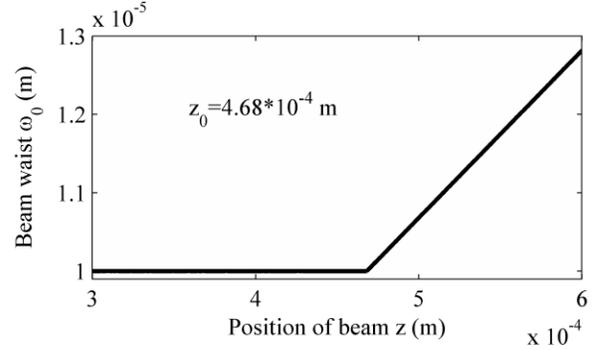

**Fig. 2(a)** Estimated beam waist from equation 4 with respect to position.

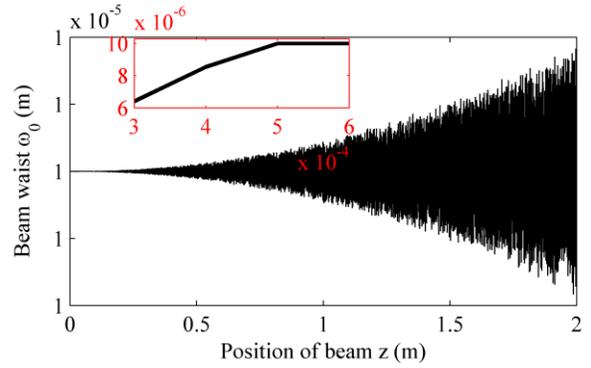

**Fig. 2(b)** Estimated beam waist from equation 5 with respect to position.

Fig. 1 depicts the behavior of inequality (Eq. 6) at different positions. As shown in the inset, at Rayleigh position ($z=z_0$) the inequality became equal to zero and at this position both the Eqs. 4 and 5 give the same solution for Eq. 3. On either side of Rayleigh position, either one of the Eqs. 4 and 5 gives physically acceptable solution while another one gives the non-physical solution.

### 3. Verification of equations

#### 3.1 Theoretical verification

To find physically acceptable solutions of Eq. 3 over the range of beam position, we have chosen 10 μm beam waist with wavelength 671 nm and the corresponding Rayleigh length is 468 μm. At each position of the beam, we found the beam spot size by Eq. 1 and then re-estimated these values in the equations Eqs. 4 and 5. The obtained beam waist data from Eqs. 4 and 5 are shown in Figs. 2(a) and (b). From Fig. 2(a), Eq. 4 gives the physically acceptable solution for Eq. 3 when $z \leq z_0$ (we got same beam waist what we initially we have chosen for this calculation) and for $z>z_0$ the beam waist increasing with increasing laser beam position with respect to beam waist. Similarly for $z \geq z_0$, as depicted in the Fig. 2(b), Eq. 5 gives the physically acceptable beam waist with the minor error. At 2 m position gives the error in the spot size around $10^{-15}$ m which is very small as compared to the beam waist (10 μm). Generally, spot sizes present greater than micro-meter due to the diffraction limit of the beam waist. Thus its effect on the spot size measurement can be neglected. As shown in the inset of the Fig. 2(b), this solution becomes unphysical and hence not acceptable for the case of $z<z_0$.

### 3. Experimental

#### 3.2 Experimental verification

To experimentally validate this method, we have carried out the pinhole experiment with 632 nm wavelength He-Ne laser. We have used 0.5 mm pinhole with 0.1 μm spatial resolution New Port XPS translation stage for scanning the beam waist. The transmitted signal from pinhole was measured by an optical power meter (Model 842-PE from New Port). The pinhole and optical detector are optically aligned on a translation stage to avoid the transverse effects while scanning the beam. Fig. 3 projects the experimental setup of the spot size measurement. The first lens ($L_1$), we have used to maximize the spot size and second ($L_2$) lens for experimental spot size verification.

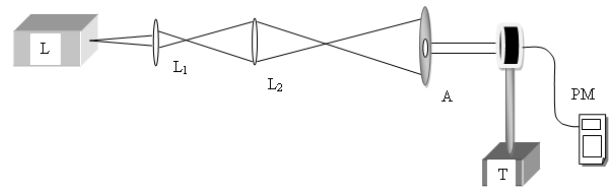

**Fig. 3** Experimental setup of spot size measurement: L→ He-Ne Laser, $L_1$→ First lens, $L_2$→ Second lens,

A→ Pinhole, PM→ Power meter, T→ XPS Translation stage.

As seen from Fig. 4(a), from the pinhole scan of the beam profile, we have estimated 4.7 cm as spot size at 31.7 cm from the beam waist. By using Eq. 5, we obtained $1.36 \times 10^{-2}$ mm as waist size at the focus of the second lens. With help of Eq. 1, we estimated spot-size of 4.29 mm at the second lens position and within the error bar, it is equal to the spot size at the lens 4.3 mm, as estimated from the pinhole scan (Fig. 4(b)).

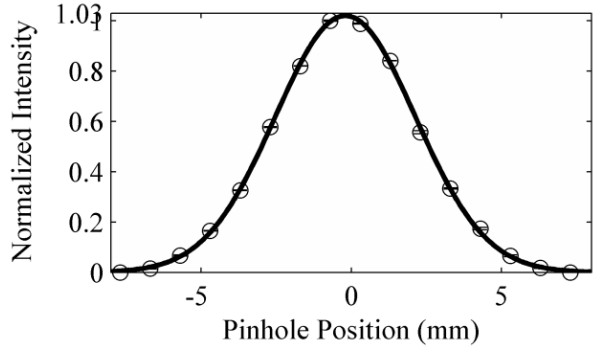

**Fig. 4(a)** Pinhole scanned laser beam profile at z=31.7±0.1 cm from the beam waist of the second lens.

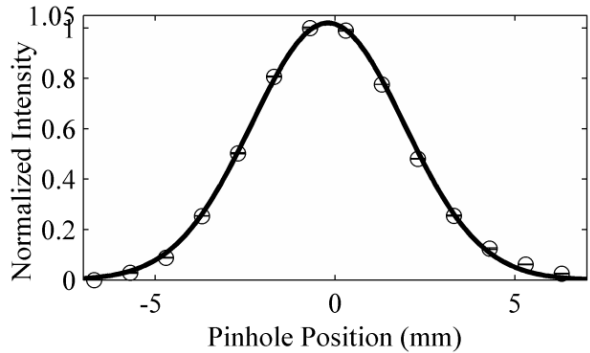

**Fig. 4(b)** Pinhole scanned laser beam profile at second lens position.

### 3. Conclusion

We have derived equations for beam waist as a function of variable spot size from well-known Gaussian beam profile equations. With these equations, a method for indirect measurement of the smaller spot sizes (<1 mm) of high peak power lasers with affordable optical components (Pinhole or knife edge) has been expounded. Equations derived from well physical meaning equations are not necessarily physically valid. In this paper, we have shown how the physical acceptable mathematical procedure equation deviates from their original physical meaning while in the mathematical analysis process. So care must be taken while we are using new equations in the physical interpretation even though they have been derived from physically acceptable equations. It is a facilitative and good technique to use the optical detectors without thermal damage.

**Acknowledgment:** Acknowledgements are to CIF of Pondicherry University for providing experimental facilities.